\documentclass[arxiv, usenatbib]{iupartex-v22}

\usepackage{newtxtext,newtxmath}
\usepackage[T1]{fontenc}
\usepackage{mathtools}
\usepackage{physics}
\usepackage{listings}

\usepackage{bm}

\bibpunct{(}{)}{;}{a}{}{,}
\usepackage{url}
\usepackage{ulem} 
\usepackage{subcaption}
\usepackage{minted} 

\newcommand{\<}{\begin{equation}}
\newcommand{\?}{\end{equation}}

\title[Analysis of YSOs]{Parametric SED Modelling of Protoplanetary Discs: Validation and Application to an Unstudied YSO}

\author[Bak{\i}\c{s} \& Habal{\i}]{%
V. Bak{\i}\c{s}$^{1}$,\orcid{0000-0002-3125-9010}
and
A. Y. Habal{\i} $^{2}$\orcid{0000-0001-5411-9654}
\affsep \\
$^1$Akdeniz University, Faculty of Science, Department of Space Sciences \& Technologies, 07058, Antalya, T\"{u}rkiye\\
$^2$Akdeniz University, Serik G\"{u}ls\"{u}n-S\"{u}leyman S\"{u}ral Vocational School, Department of Medical Services and Techniques,\\~~~Opticianry Program,  Antalya, T\"{u}rkiye
}

\corres{Volkan Bak{\i}\c{s}}{volkanbakis@akdeniz.edu.tr}

\pubyear{2026}

\doiheader{XXXXXXX/PAR.2025.00000}
\date{
	\pSubmit{XX.XX.2026} 
	\pRevReq{XX.XX.2026}
	\pLastRevRec{XX.XX.2026}
	\pAccept{XX.XX.2026}
	\pPubOnl{XX.XX.2026}
}
\volume{4}
\volnumber{1}

\begin{document}
\label{firstpage}
\pagerange{\pageref*{firstpage}--\pageref*{lastpage}}
\maketitle

\begin{abstract}
We present a physically motivated spectral energy distribution (SED) modelling framework for deriving stellar and circumstellar disc parameters from broadband photometry. The model combines a parametrized disc structure, dust opacity, and interstellar extinction within a Bayesian Markov Chain Monte Carlo (MCMC) inference scheme, allowing correlated parameters to be constrained self-consistently. Initial parameter estimates are obtained via non-linear least-squares fitting and subsequently refined through MCMC sampling. The method is first validated using the well-studied debris disc system 49 Cet, for which the model successfully reproduces key literature properties. It is then applied to the previously uncharacterised young stellar object (YSO) candidate 2MASS J02512618+6012576, using photometric measurements compiled from multiple surveys. The resulting fit indicates a late-type pre-main-sequence star surrounded by a substantial circumstellar disc consistent with a moderately embedded Class II object. We further assess the sensitivity of the inferred parameters to the adopted extinction law and find that the high reddening required by the model is robust against variations in $R_V$. This work demonstrates that physically meaningful constraints on disc structure can be obtained from broadband SED modelling when extinction and distance are treated within a statistically consistent framework.
\end{abstract}




\section{Introduction} 
\par
The formation of stars and planetary systems is one of the most fundamental processes in modern astrophysics. Stars like our Sun originate from the gravitational collapse of dense molecular cloud cores, leading to the formation of a central YSO surrounded by a rotating, flared accretion disc and a larger-scale circumstellar envelope \citep{Lada1987}. Understanding the physical and chemical evolution of these environments is crucial for constraining the initial conditions of planet formation.

Over the past few decades, the study of circumstellar environments has been revolutionized by high-resolution observations across the electromagnetic spectrum. Space missions such as the \textit{Spitzer Space Telescope} \citep{Spitzer2004} and the \textit{Herschel Space Observatory} \citep{Herschel2010} have provided extensive catalogues of SEDs, revealing infrared excesses that characterize dust emission from discs and envelopes. More recently, interferometric facilities like the \textit{Atacama Large Millimeter/submillimeter Array} (ALMA) \citep{Alma2009} and adaptive optics instruments like \textit{VLT/SPHERE} \citep{Sphere2019} have enabled us to resolve the substructures within these discs, such as gaps, rings, and spirals, at astronomical unit scales.

Interpreting these observations requires sophisticated radiative transfer modelling. The interaction between stellar radiation and circumstellar dust governs the thermal structure and the observed emission of the system. Historically, analytical models such as those by \cite{Chiang1997} established the foundation for understanding flared, passive discs in radiative equilibrium. However, the complex interplay between the shadowed midplane and the irradiated surface layers often necessitates numerical solutions.

While comprehensive Monte Carlo radiative transfer codes like {\sc radmc-3d} \citep{Dullemond2012} and {\sc mcfost} \citep{Pinte2006} offer high precision, they are computationally intensive. There remains a significant need for efficient, ray-tracing-based models that can rapidly explore the parameter space and provide synthetic observables (SEDs and images) to compare with multi-wavelength data.

In this work, a numerical model designed to simulate the emission from a YSO-disc-envelope system is presented. By combining a power-law density distribution for the disc with a 3D line-of-sight integration for the spherical envelope, our model captures the transition from optically thick to thin regimes across a wide frequency range. This tool allows for the rapid synthesis of multi-wavelength images and SEDs, facilitating a deeper understanding of how geometry and inclination affect the observed properties of young stars.

\section{Methodology}
\subsection{Stellar and Dust Input Physics}
\subsubsection{Stellar Photosphere}

The stellar contribution to the SED is modelled assuming a centrally located star that dominates the radiation field at short wavelengths. The stellar photosphere is represented by a blackbody spectrum characterized by the effective temperature $T_\star$ and stellar radius $R_\star$. The intrinsic stellar flux density at wavelength $\lambda$ is expressed as

\begin{equation}
F_{\nu,\star}(\lambda) = \left( \frac{R_\star}{d} \right)^2 \pi B_\nu(\lambda, T_\star),
\end{equation}
where $d$ is the distance to the system and $B_\nu$ denotes the Planck function (e.g., standard radiative processes). This simplified representation is sufficient for disc-dominated SED modelling, particularly at optical and near-infrared wavelengths, where stellar emission provides the primary heating source for the circumstellar material.

The stellar radiation field serves as the main driver of passive disc heating in the model. Effects such as stellar spots, accretion luminosity, or chromospheric emission are not explicitly included, and the star is assumed to radiate isotropically. These assumptions are appropriate for debris discs and passively irradiated circumstellar discs, where accretion-related excess emission is expected to be negligible.

\subsubsection{Dust Opacity and Extinction}

The interaction between radiation and circumstellar dust is governed by wavelength-dependent dust opacities, which determine both the thermal emission and the attenuation of radiation along the line of sight. In the present model, dust grains are assumed to be in thermal equilibrium and to emit as modified blackbodies, with the opacity implicitly encoded in the disc emission prescription.

Interstellar and circumstellar extinction are incorporated through a wavelength-dependent extinction law parameterised by the colour excess $E(B-V)$ and the total-to-selective extinction ratio $R_{\rm V}$. The extinction at each wavelength is applied to the intrinsic model SED as

\begin{equation}
F_{\nu,\mathrm{obs}}(\lambda) = F_{\nu,\mathrm{int}}(\lambda) \times 10^{-0.4 A_\lambda},
\label{eq:Av}
\end{equation}
where $A_\lambda$ is the extinction at wavelength $\lambda$ which is computed using the extinction law of \citet{Gordon2023}. The extinction curve is computed using a standard parametrization appropriate for the diffuse interstellar medium, while $E(B-V)$ is treated as a free parameter during the SED fitting process.

This approach allows the model to self-consistently account for reddening effects without introducing additional degeneracy in the disc emission parameters. The adopted dust treatment is intentionally simplified, as the primary aim of the model is to capture the global structure of the SED rather than detailed grain-size distributions or compositional effects.

\subsection{Circumstellar disc Model}

The circumstellar disc is modelled as an axisymmetric, geometrically thin but vertically flared structure surrounding the central star. The disc is assumed to be passively heated by stellar irradiation, with no internal viscous heating included. The physical structure of the disc is described using simple parametric prescriptions that allow efficient exploration of the parameter space during SED fitting.

\subsubsection{Surface Density Profile}

The radial distribution of disc material is described by a power-law surface density profile (e.g., \citealt{LyndenBell1974}) of the form

\begin{equation}
\Sigma(r) = \Sigma_0 \left( \frac{r}{r_0} \right)^{-p},
\end{equation}
where $\Sigma_0$ is the surface density normalization at a reference radius $r_0$, and $p$ is the surface density exponent. The disc extends from an inner radius $R_{\mathrm{in}}$ to an outer radius $R_{\mathrm{out}}$.

This parametrization is commonly adopted in circumstellar disc studies and provides a flexible yet physically motivated description of the radial mass distribution. In the present model, $\Sigma_0$ is treated as a free parameter, while the exponent $p$ may be fixed or varied depending on the fitting strategy.

\subsubsection{Vertical Structure and Flaring}

The vertical structure of the disc is described by a Gaussian density distribution, assuming hydrostatic equilibrium in the vertical direction. The scale height $H(r)$ follows a flared geometry (e.g., \citealt{Chiang1997,Dullemond2001}) given by

\begin{equation}
H(r) = H_0 \left( \frac{r}{r_0} \right)^{\beta},
\end{equation}
where $H_0$ is the scale height at the reference radius $r_0$, and $\beta$ is the flaring index. This prescription captures the increasing vertical thickness of the disc at larger radii due to stellar irradiation.

The flaring index $\beta$ plays a critical role in shaping the mid- to far-infrared SED, as it controls the amount of stellar radiation intercepted by the disc surface. Larger values of $\beta$ correspond to more strongly flared discs, resulting in enhanced infrared emission at long wavelengths.

\subsubsection{Thermal Structure: Passive Irradiation}

The disc temperature structure is computed assuming passive heating by stellar irradiation. The local dust temperature is approximated by a power-law dependence on radius (e.g., \citealt{Chiang1997, Dullemond2001}),

\begin{equation}
T(r) = T_0 \left( \frac{r}{r_0} \right)^{-q},
\end{equation}
where $T_0$ is the temperature at the reference radius $r_0$, and $q$ is the temperature gradient exponent. Typical values of $q$ range between 0.5 and 0.75 for passively irradiated discs. This formulation reflects radiative equilibrium between absorbed stellar radiation and thermal emission from dust grains.

\begin{figure}[t]
    \centering
    \includegraphics[width=0.75\textwidth]{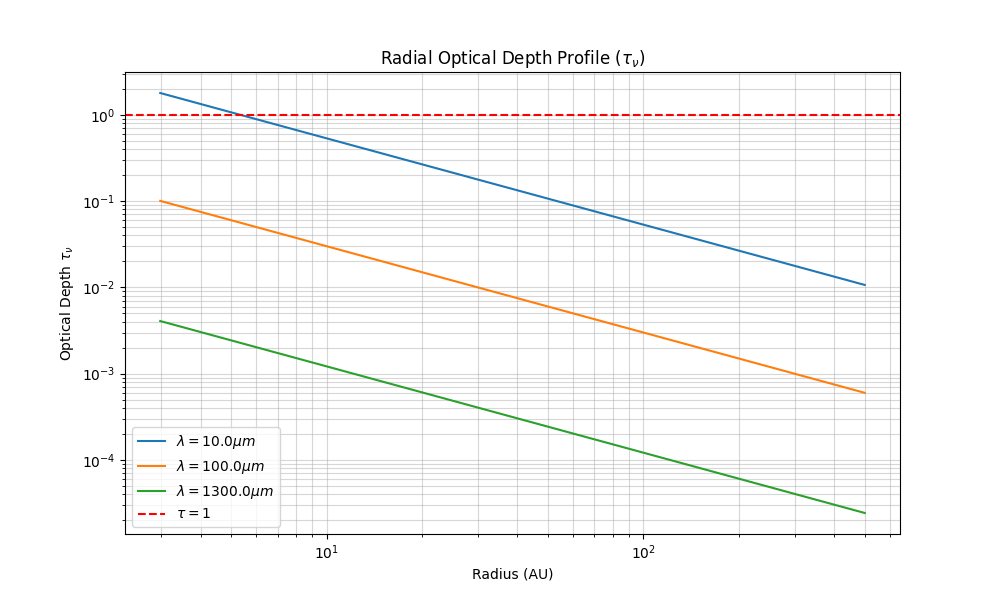}
	\caption{
    Radial optical depth profiles of the circumstellar disc for 2MASS J02512618+6012576 (in Section~4) at $\lambda = 10$, 100, and 1300 $\mu$m. The horizontal dashed line marks $\tau_\nu = 1$, corresponding to the transition between optically thick ($\tau_\nu > 1$) and optically thin ($\tau_\nu < 1$) regimes. The radial location of this transition varies with wavelength, reflecting the wavelength dependence of the dust opacity and the disc surface density structure.}
    \label{fig:tau_radius}
\end{figure}

The efficiency of stellar irradiation in heating the disc is governed by the wavelength-dependent optical depth, which controls the penetration of stellar photons into the disc structure. Figure~\ref{fig:tau_radius} shows the radial optical depth profiles computed at $\lambda = 10$, $100$, and $1300\,\mu$m. At mid-infrared wavelengths, the inner disc remains optically thick, enabling efficient absorption and reprocessing of stellar
radiation. In contrast, the disc becomes largely optically thin at millimeter wavelengths, indicating that long-wavelength emission primarily traces the bulk dust mass rather than the thermally heated surface layers.

The passive irradiation assumption is appropriate for discs with low or negligible accretion rates, such as debris discs or evolved protoplanetary discs. Viscous heating, self-shadowing, and full radiative transfer effects are not explicitly included, and the resulting temperature structure should therefore be regarded as an effective description rather than a self-consistent solution.

\subsubsection{Disc SED Calculation}

The disc SED is computed by integrating the thermal emission from concentric annuli over the full radial extent of the disc. Each annulus emits as a modified blackbody at the local temperature $T(r)$, weighted by the surface density and geometrical projection effects.

The total disc flux density is computed by integrating the emission from concentric annuli (e.g., \citealt{Beckwith1990, Andrews2009}),

\begin{equation}
F_{\nu,\mathrm{disc}} = \frac{1}{d^2} \int_{R_{\mathrm{in}}}^{R_{\mathrm{out}}}
2\pi r \, B_\nu[T(r)] \, \left( 1 - e^{-\tau_\nu(r)} \right) \, dr,
\end{equation}
where $d$ is the distance to the system and $\tau_\nu(r)$ is the effective optical depth. In the optically thin regime, this expression reduces to a linear dependence on the surface density, while optically thick regions naturally saturate the emission.

This semi-analytic approach enables rapid computation of disc SEDs while retaining the essential physical dependencies on disc geometry, temperature structure, and dust content.

\subsection{Circumstellar Envelope Model}

In addition to the circumstellar disc, an extended circumstellar envelope can be optionally included in the model to account for excess emission at mid- and far-infrared wavelengths. The envelope component is primarily intended to represent remnant material in young or embedded systems and is not required for more evolved or debris disc systems.

The inclusion of the envelope is controlled by a logical switch, allowing the SED fitting procedure to assess whether an envelope contribution is necessary to reproduce the observed emission.

\subsubsection{Density and Temperature Distributions}

The circumstellar envelope is modeled as a spherically symmetric structure characterized by a radially decreasing density profile (e.g., \citealt{Shu1977, Adams1987}),

\begin{equation}
\rho(r) = \rho_0 \left( \frac{r}{r_0} \right)^{-p_{\mathrm{env}}},
\end{equation}
where $\rho_0$ is the density normalization at a reference radius $r_0$, and $p_{\mathrm{env}}$ is the envelope density exponent. The envelope extends from an inner radius, typically comparable to the disc outer radius, to a large outer radius encompassing the extended circumstellar environment.

The temperature structure of the envelope is described by a power-law profile (e.g., \citealt{Adams1987}),

\begin{equation}
T_{\mathrm{env}}(r) = T_{\mathrm{env},0} \left( \frac{r}{r_0} \right)^{-q_{\mathrm{env}}},
\end{equation}
where $T_{\mathrm{env},0}$ is the temperature at the reference radius and $q_{\mathrm{env}}$ controls the radial temperature gradient. This parametrization provides an effective description of radiative heating by the central star without performing full radiative transfer calculations.

\subsubsection{Envelope Emission}

The thermal emission from the envelope is computed assuming optically thin dust emission (e.g., \citealt{Hildebrand1983, Beckwith1990}) by integrating over the envelope volume,

\begin{equation}
F_{\nu,\mathrm{env}} = \frac{1}{d^2} \int_{V}
\kappa_\nu \, \rho(r) \, B_\nu[T_{\mathrm{env}}(r)] \, dV,
\end{equation}
where $d$ is the distance to the source and $\kappa_\nu$ is the dust opacity, which is parameterised as a power-law function of frequency,

\begin{equation}
\kappa_\nu = \kappa_0 \left( \frac{\nu}{\nu_0} \right)^{\beta},
\end{equation}
where $\kappa_0$ is the opacity at a reference frequency $\nu_0$, and $\beta$ is the opacity index, which reflects the grain size distribution and composition (e.g., \citealt{Beckwith1990}).

The assumption of optically thin emission is appropriate for low-density envelopes and simplifies the computation while retaining the essential dependence on dust mass and temperature. In the total SED model, the envelope contribution is added linearly to the stellar and disc components. The envelope is primarily constrained by long-wavelength data and serves to reproduce broad infrared excesses that cannot be accounted for by the disc alone.

\subsection{Radiative Transfer and Image Synthesis}

The radiative transfer calculations in this study are based on a semi-analytical approach, designed to efficiently compute SEDs and synthetic images while retaining the dominant physical dependencies on temperature, density, and dust opacity. Full multi-dimensional Monte Carlo radiative transfer is not performed; instead, the model emphasizes transparency, computational efficiency, and parameter interpretability.

\subsubsection{Total Spectral Energy Distribution}

The total SED of the system is constructed as the linear superposition of individual emission components originating from the central star, circumstellar disc, inner rim, and the circumstellar envelope when enabled.

\begin{equation}
F_{\nu}^{\mathrm{tot}} =
F_{\nu}^{\star} +
F_{\nu}^{\mathrm{disc}} +
F_{\nu}^{\mathrm{rim}} +
F_{\nu}^{\mathrm{env}}.
\end{equation}

The inner rim contribution, $F_{\nu}^\mathrm{rim}$, represents the thermal emission from the dust sublimation front at the inner edge of the disc. This component is approximated as a localized emission region at $R_{\mathrm{in}}$, where dust is heated to near-sublimation temperatures.

Each component is computed independently based on its respective physical structure and temperature distribution. The disc and envelope fluxes are obtained by integrating the local dust emission over their spatial extent, assuming axisymmetric geometry and isotropic emission.

This additive formulation allows for flexible inclusion or exclusion of individual components and facilitates direct interpretation of their relative contributions across different wavelength regimes.

\subsubsection{Optical Depth Integration and Extinction}

Line-of-sight extinction effects are incorporated by applying a wavelength-dependent attenuation factor to the intrinsic SED. The total observed flux density is expressed as

\begin{equation}
F_{\nu}^{\mathrm{obs}} = F_{\nu}^{\mathrm{tot}} \, e^{-\tau_\nu},
\end{equation}
where $\tau_\nu$ is the optical depth along the observer's line of sight, computed from the adopted extinction law \citep{Gordon2023}. The optical depth includes contributions from interstellar extinction and, where applicable, circumstellar material. Therefore, the extinction can be written equivalently either in terms of optical depth, $F_\nu^{\mathrm{obs}} = F_\nu^{\mathrm{tot}} e^{-\tau_\nu}$, or in magnitudes, $F_\nu^{\mathrm{obs}} = F_\nu^{\mathrm{tot}} 10^{-0.4A_\lambda}$, with $A_\lambda = 1.086\,\tau_\nu$.

Interstellar extinction is parameterised using the colour excess $E(B-V)$ and total-to-selective extinction ratio $R_{\rm V}$, adopting a standard Galactic extinction law. Circumstellar extinction is computed by integrating the dust opacity and density along the line of sight, assuming axisymmetric and neglecting scattering effects.

This treatment captures the dominant attenuation behaviour while avoiding the computational complexity of full radiative transfer calculations.

\subsubsection{Multi-wavelength Image Synthesis}

Synthetic images are generated by computing the spatially resolved specific intensity distribution of the disc at selected wavelengths. The disc is projected onto the plane of the sky for a given inclination angle, and the emergent intensity is calculated as a function of position, assuming optically thin or moderately thick emission.

For each wavelength, the intensity map is constructed by integrating the local thermal emission along the line of sight, assuming optically thin dust emission (e.g., \citealt{Hildebrand1983}),

\begin{equation}
I_\nu(x, y) = \int \kappa_\nu \, \rho(s) \, B_\nu[T(s)] \, ds,
\end{equation}
where $s$ denotes the path length through the disc. The resulting images are normalized and displayed on physical spatial scales expressed in astronomical units.

Multi-wavelength image synthesis enables direct comparison of disc morphology at different infrared wavelengths and provides qualitative insight into the spatial origin of the SED features, particularly the transition from stellar-dominated to disc-dominated emission regimes.

\subsection{SED Fitting and Analysis}

\subsubsection{Observed Data Preparation}

Observed SED data are compiled using the SIMBAD database as an aggregation interface to the VizieR photometric services. The collected measurements originate from multiple surveys (e.g., {\it Gaia}; \citealt{GaiaDR3}, 2MASS; \citealt{Skrutskie2006}, {\it WISE}; \citealt{Wright2010}, Pan-STARRS; \citealt{Chambers2016}, and {\it IRAS}; \citealt{Neugebauer1984}), each characterised by its own instrumental apertures, calibration procedures, and photometric systems. Therefore, no attempt is made to impose a uniform aperture across the dataset.

An initial automatic filtering is applied to remove non-physical entries (e.g., non-finite or non-positive flux values). Subsequently, an interactive cleaning procedure is performed to discard evident outliers and unreliable measurements. No formal sigma-clipping algorithm is applied, as the number of data points per wavelength region is limited and the dataset is heterogeneous in origin, making automated rejection schemes less reliable.

Following this step, measurements at identical or nearly identical wavelengths are grouped and combined by taking the arithmetic mean of their flux values, irrespective of their survey origin. The corresponding uncertainties are propagated within each group to obtain a representative error for the merged data points.

In cases where the reported uncertainties are missing or non-physical, wavelength-dependent fractional uncertainties are adopted to ensure a consistent weighting scheme in subsequent analyses. The final cleaned SED consists of a reduced set of representative wavelength--flux pairs, which are stored locally and reused in subsequent model evaluations. This approach ensures reproducibility and avoids repeated queries from the database.

\subsubsection{Free and Fixed Parameters}

The model parameters are divided into two categories: free parameters, which are optimized during the fitting process, and fixed parameters, which are kept constant based on independent observational constraints or values from the literature.

Free parameters typically include quantities governing the disc structure and dust content, such as the surface density normalization, inner disc radius, and vertical scale height. Fixed parameters may include stellar properties (e.g., effective temperature and distance) or disc power-law indices when insufficient constraints are available from the SED alone.

This separation is implemented through a centralized parameter management system, allowing flexible selection of free and fixed parameters without modifying the underlying model equations. Each free parameter is assigned physically motivated lower and upper bounds to prevent unphysical solutions.

\subsubsection{SED Fitting Procedure}

The fitting procedure minimizes the difference between the observed and modeled SEDs using a non-linear least-squares optimization based on the \texttt{curve\_fit} algorithm from the SciPy library \citep{Virtanen2020}. The model SED is evaluated at the observed wavelengths during the optimization, while a denser wavelength grid is used for visualisation of the final best-fit solution.

At each iteration, the free parameters are updated and propagated consistently through the stellar, disc, rim, and envelope components of the model. The optimization is initialized using physically reasonable starting values and is terminated when convergence is reached, or a predefined maximum number of iterations is exceeded.

This deterministic fitting approach provides rapid convergence and is well-suited for exploratory analysis and parameter sensitivity studies.

\subsubsection{Goodness-of-Fit and Residuals}

The quality of the fit is assessed by direct comparison of the observed and modelled SEDs in logarithmic wavelength–flux space. Residuals are computed as the fractional difference between the observed and modelled flux densities,

\begin{equation}
\mathrm{Residual} = \frac{F_{\nu}^{\mathrm{obs}} - F_{\nu}^{\mathrm{model}}}{F_{\nu}^{\mathrm{obs}}}.
\end{equation}

Residuals are inspected as a function of wavelength to identify systematic deviations associated with specific spectral regions, such as near-infrared excess or long-wavelength disc emission. This qualitative assessment provides insight into which physical components dominate the mismatch between the model and observations.

\subsubsection{Parameter Degeneracies and Limitations}

SED-based disc modelling is inherently affected by parameter degeneracies, particularly among quantities that regulate the overall normalization and temperature structure of the disc emission. Variations in surface density normalization, disc scale height, and dust opacity can produce similar effects on the infrared and (sub-)millimeter flux levels, leading to non-unique solutions when relying on broadband photometry alone.

\begin{figure}[t]
    \centering
    \includegraphics[width=0.75\textwidth]{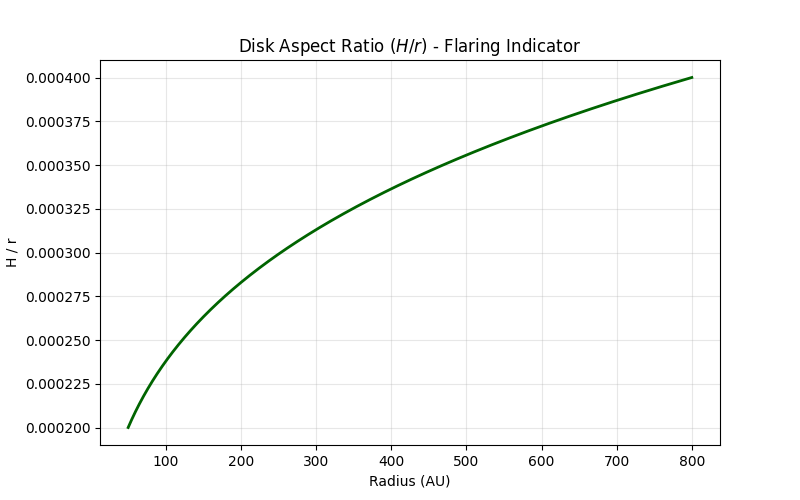}
    \caption{
    Radial profile of the disc aspect ratio ($H/r$) adopted in the circumstellar disc model. The vertical structure is parameterised as a power-law function of radius, $H(r) = H_0 (r / R_{\mathrm{in}})^{\beta_{\rm H}}$, where $H_0$ is the scale height at the inner disc radius $R_{\mathrm{in}}$ and $\beta_{\rm H}$ controls the disc flaring. This prescription governs the irradiation geometry and thus directly affects the resulting SED.
    }
    \label{fig:aspect_ratio}
\end{figure}

Geometric parameters introduce additional sources of uncertainty in SED-based modelling. In particular, the disc aspect ratio ($H/r$) plays a key role in shaping the infrared excess by regulating the degree of disc flaring and the efficiency of stellar irradiation intercepted by the disc surface.  Figure~\ref{fig:aspect_ratio} illustrates the adopted radial profile of $H/r$ used in the present model, which defines the disc geometry and thermal structure. However, SED data alone provide limited constraints on the vertical disc structure, and the inferred flaring properties should therefore be interpreted as model-dependent rather than uniquely determined by the observations.

These degeneracies are further coupled with inclination and inner disc radius, which remain only weakly constrained in the absence of spatially resolved observations. As a result, the derived parameter values should be interpreted as representative rather than unique physical solutions.

The present fitting framework prioritizes physical transparency and computational efficiency over an exhaustive exploration of the parameter space. While advanced techniques such as MCMC sampling could provide a more comprehensive characterization of uncertainties and parameter correlations, such approaches are beyond the scope of this exploratory study.

\section{Application to a Well-Studied System: 49 Cet}

To validate the modelling framework and fitting methodology, we apply our SED model to the well-studied debris disc system 49 Cet. This source is a nearby A-type star hosting a gas-rich but dust-poor debris disc, making it an ideal benchmark for testing simplified disc models without invoking complex envelope structures.

\subsection{Stellar and Observational Properties}

49 Cet is an A1V star located at a distance of approximately 57~pc \citep[][]{GaiaDR3}. Its stellar parameters, including effective temperature and radius, are well constrained in the literature, allowing these quantities to be fixed during the SED fitting process. The system exhibits a clear infrared excess at mid- and far-infrared wavelengths, attributed to thermal emission from circumstellar dust, while showing minimal excess at near-infrared wavelengths.

The observed SED data are compiled from the SIMBAD database and subsequently cleaned and averaged following the procedure described in Section~2.5.1. The resulting dataset spans wavelengths from the optical to the millimeter regime, providing strong constraints on the dust temperature distribution and disc mass.

\subsection{Model Setup and Assumptions}

Given the debris disc nature of 49 Cet, the circumstellar environment is modelled solely with a passive dust disc component. The circumstellar envelope is disabled for this system, consistent with the absence of observational evidence for a remnant protostellar envelope.

The disc is assumed to be geometrically thin and optically thin at most wavelengths, with a power-law surface density profile and a passive temperature structure governed by stellar irradiation. An inner disc cavity is introduced to reproduce the lack of near-infrared excess, while the outer disc radius is chosen sufficiently large to encompass the cold dust responsible for far-infrared and sub-millimeter emission.

\subsection{SED Fitting Results}

The SED fitting is performed using a non-linear least-squares optimization, with selected disc parameters treated as free variables. These typically include the surface density normalization, inner disc radius, and vertical scale height, while stellar parameters and distance are fixed.

\begin{figure}[ht]
    \centering
    \includegraphics[width=0.85\linewidth]{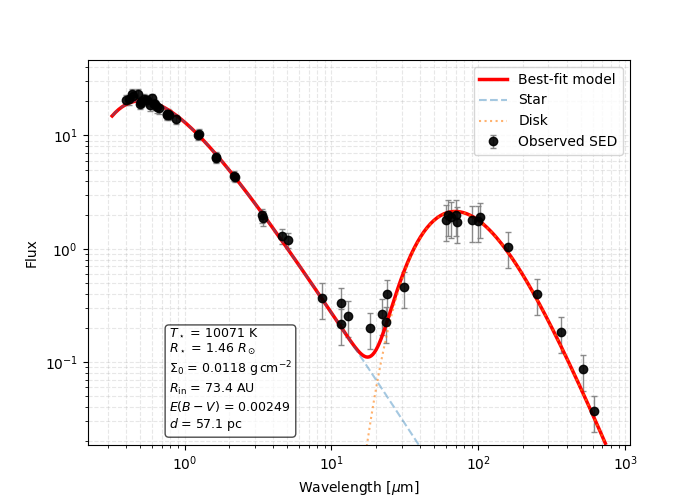}
    \caption{Observed and modeled SED of 49 Cet. Individual contributions from the stellar photosphere and disc emission are shown, along with the total extincted model SED.}
    \label{fig:49cet_sed_fit}
\end{figure}

Figure~\ref{fig:49cet_sed_fit} presents the observed SED of 49 Cet together with the best-fit model. The model successfully reproduces the overall shape and normalization of the infrared excess, particularly at wavelengths longer than $\sim10~\mu$m. The lack of significant excess at shorter wavelengths is naturally explained by the presence of an inner disc cavity.

\subsection{disc Structure and Physical Interpretation}

The fitted model indicates that the dust responsible for the observed emission is predominantly cold, with characteristic temperatures corresponding to disc radii of several tens of astronomical units. The inferred surface density normalization implies a low total dust mass, consistent with the debris disc classification of the system.

The total dust mass, $M_{\mathrm{dust}}$, is obtained by integrating the surface density profile over the disc,

\begin{equation}
M_{\mathrm{dust}} = \int_{R_{\mathrm{in}}}^{R_{\mathrm{out}}} 2\pi r \, \Sigma(r)\, dr.
\end{equation}
This quantity represents the dust mass only and does not include any gas contribution. A gas-to-dust ratio is not assumed in this work.

The absence of a significant near-infrared excess suggests that hot dust is either absent or present only at very low levels, supporting the interpretation of an evolved disc with substantial inner clearing. While the model reproduces the global SED well, degeneracies between disc mass, dust opacity, and vertical structure remain, as discussed in Section~2.5.5.

\subsection{Comparison with Literature}

The derived disc properties for 49 Cet are broadly consistent with previous studies. Literature estimates classify 49 Cet as a debris disc with a low dust mass (e.g., $M_{\rm dust} \sim 0.3~M_\oplus$, \citealt{Nhung2017}) and a cleared inner cavity of $\sim$26-73~AU \citep{Hughes2017}. Our simplified SED-based modelling yields a comparable dust mass of $M_{\rm dust} \sim 0.128~M_\oplus$ and an inner radius (75.5$^{13}_{10}$ AU consistent with this range, despite not explicitly including scattered light or gas emission. While more sophisticated radiative transfer models would capture additional features such as scattering asymmetries or gas line emission, the present results reproduce the essential characteristics of the disc. Applying this approach to a well-characterized system like 49 Cet validates the internal consistency of the modelling framework and establishes a baseline for subsequent application to less studied or newly discovered YSOs.

\section{Application to a Previously Unstudied System}

In this section, the modelling framework developed in this study is applied to a circumstellar system for which no detailed SED modelling has previously been reported. The purpose is twofold: (i) to test the robustness of the method when applied to a poorly characterized source, and (ii) to evaluate which physical constraints can be reliably derived using only broadband photometric data.

The target, 2MASS J02512618+6012576, was selected from the YSO catalog of \citet{Habali2026}. In the SIMBAD database, the object is classified as a Young Stellar Object (YSO) candidate. According to {\it Gaia} DR3 \citep{GaiaDR3}, the source has an apparent magnitude of $m_{\rm G} = 19.326 \pm 0.004$ and a trigonometric parallax of $\varpi = 0.9448 \pm 0.2435$ mas, corresponding to a nominal distance of $d \approx 1059$ pc. The object lies within the projected boundaries of IC~1848 in the Soul Nebula \citep{Sharpless1955}. To our knowledge, no dedicated SED-based physical characterization of this source has been published. However, it is listed in the recent catalogue of YSOs by \cite{Habali2026} as a Class II object.

A total of 28 photometric measurements covering the wavelength range 4400--22000\,\AA\ were compiled for the analysis. While the data were accessed via the SIMBAD database, the original measurements are drawn from multiple surveys, including {\it Gaia} \citep{GaiaDR3}, 2MASS \citep{Skrutskie2006}, {\it WISE} \citep{Wright2010}, and Pan-STARRS \citep{Chambers2016}, each providing fluxes based on their own instrumental apertures and calibration procedures. Therefore, no uniform aperture was applied across all datasets; instead, the reported catalog fluxes are adopted as representative of the source emission. Given the absence of nearby bright sources within a few arcseconds, the measurements are assumed to consistently trace the target flux across different surveys. SIMBAD is used solely as an aggregation interface to identify and retrieve the corresponding measurements.

Reported uncertainties span $10^{-7}$ to $6 \times 10^{-4}$ Jy. The {\it Gaia} DR3 catalog \citep{GaiaDR3} lists a colour index of $G_{\mathrm{BP}} - G_{\mathrm{RP}} = 2.9648$ mag, while a direct computation from the tabulated fluxes yields $G_{\mathrm{BP}} - G_{\mathrm{RP}} = 2.37 \pm 0.07$ mag. If interpreted without extinction, such colours would imply a relatively cool photosphere. However, given the location of the source within a dense star-forming environment, reddening is expected to significantly affect the observed optical colours.

For this reason, the SED fitting is initiated with relatively low stellar effective temperatures, allowing the colour excess $E(B-V)$ to be constrained directly by the fit. The resulting $E(B-V)$ is then used to estimate extinction-corrected Gaia colours using the extinction coefficients given by \citet{Bakis2022},

\[
E(G_{\mathrm{BP}} - G_{\mathrm{RP}}) = 1.609\,E(B-V).
\]
This procedure enables a self-consistent refinement of the stellar temperature estimate within the Bayesian framework.

The {\it Gaia} distance is incorporated as a Gaussian prior centered at $d = 1059$ pc with $\sigma = 273$ pc. Allowing the distance to vary freely would introduce a strong degeneracy with stellar radius and extinction, as the observed flux scales with $(R_\star/d)^2$ and is further modulated by reddening. Constraining the distance within observational uncertainties significantly stabilizes the parameter inference.

The infrared excess observed in the SED clearly indicates the presence of a circumstellar disc. The adopted disc model is characterized by the inner radius ($R_{\mathrm{in}}$), outer radius ($R_{\mathrm{out}}$), inclination ($i$), scale height at the inner radius ($H_0$), flaring index ($\beta_{\rm H}$), surface density normalization ($\Sigma_0$), surface density exponent ($p_{\Sigma}$), reference opacity ($\kappa_0$), and opacity index ($\beta$). Interstellar extinction is parameterised by $E(B-V)$ and the total-to-selective extinction ratio $R_{\rm V}$.

An initial estimate of the model parameters was obtained using a non-linear least-squares fitting procedure implemented via \texttt{curve\_fit}. These best-fit values were then used as starting points for an MCMC analysis performed with the \texttt{emcee} package \citep{ForemanMackey2013}. The analysis employed 20 walkers and 5000 steps per walker, with initial walker positions distributed in a small Gaussian ball around the least-squares solution.

During the MCMC analysis, the free parameters are $T_\star$, $R_\star$, $R_{\mathrm{in}}$, $\Sigma_0$, and $E(B-V)$. The distance is constrained by the {\it Gaia} prior, while $i$, $H_0$, $p_{\Sigma}$, and $\beta$ are fixed within physically motivated ranges to reduce parameter degeneracy and ensure numerical stability.

To ensure convergence, the initial portion of each chain was discarded, and only the remaining samples were used to construct the posterior distributions. Convergence was assessed by visual inspection of the chains and by verifying the stability of the posterior distributions. The posterior distributions are approximately unimodal and close to Gaussian for most parameters. The acceptance fraction was found to be $\sim 0.38$, consistent with efficient sampling.

\begin{figure*}
\centering
 \includegraphics[width=0.95\linewidth]{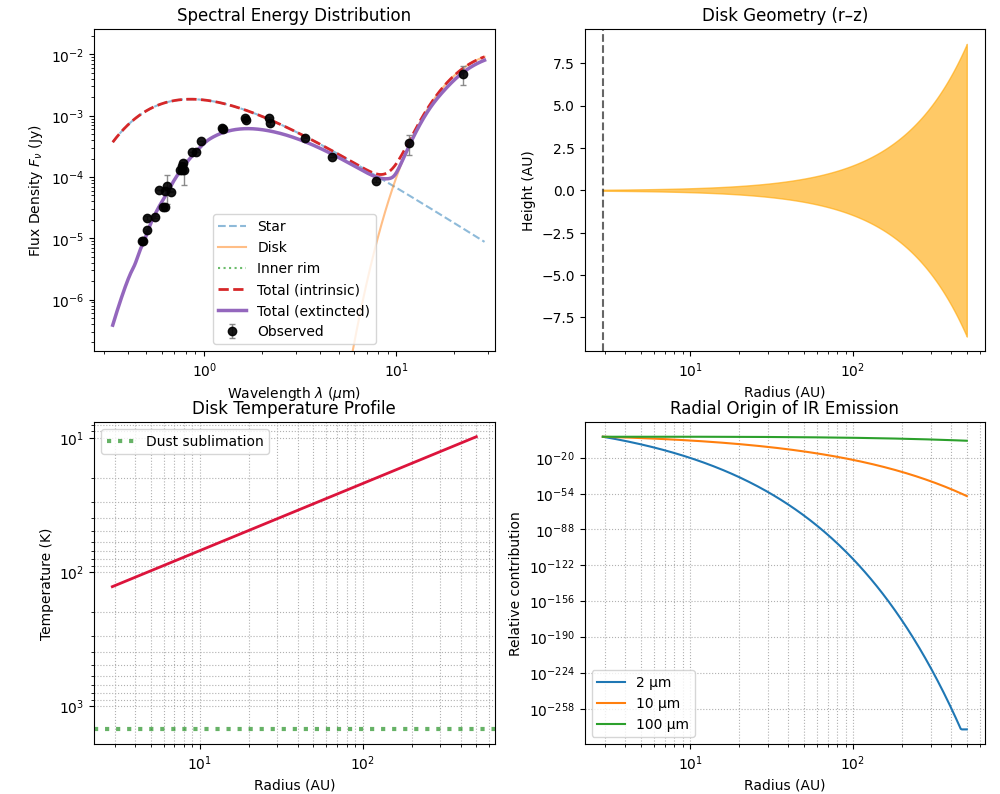}
\caption{modelling results for 2MASS J02512618+6012576. (\emph{top left}) Observed SED and best-fit model, showing the total extincted flux together with stellar and disc components. (\emph{top right}) Adopted disc geometry based on posterior median parameters. (\emph{bottom left}) Radial temperature profile of the disc midplane. (\emph{bottom right}) Radial contribution to the infrared emission, illustrating the dominant emitting regions at different wavelengths.}
\label{fig:disc_model}
\end{figure*}

The modelling results are summarised in Figure~\ref{fig:disc_model}, which illustrates both the observational fit and the physical structure of the inferred disc.

Panel (\emph{top left}) of Figure~\ref{fig:disc_model} shows the observed photometric data together with the best-fit SED model. The total extincted flux (solid curve) successfully reproduces the optical slope and the infrared excess. The stellar photosphere and disc contributions are shown separately, highlighting the dominance of the disc emission at longer wavelengths.

Panel (\emph{top right}) of Figure~\ref{fig:disc_model} presents a schematic representation of the adopted disc geometry using the posterior median parameters. The disc extends from $R_{\mathrm{in}} \approx 2$ AU to $R_{\mathrm{out}} = 500$ AU with a moderate flaring index. The relatively large inner radius is notable when compared to the expected dust sublimation radius. Dust sublimation refers to the phase transition in which solid dust grains are thermally destroyed and converted directly into gas when their temperature exceeds a critical sublimation threshold. This process defines the innermost radius at which dust can survive in circumstellar discs, commonly referred to as the dust sublimation radius. For typical silicate grains, dust sublimates at temperatures of $T_{\rm sub} \sim 1400$--$1800$ K (e.g., \citealt{Pollack1994,Kobayashi2011}), which corresponds to a characteristic radius of $R_{\rm sub} \sim 0.03$--$0.1$ AU \citep{Dullemond2001} for a late-type pre-main-sequence star. The inferred value of $R_{\mathrm{in}}$ is therefore significantly larger than the nominal sublimation radius.

This discrepancy may indicate either the presence of an inner disc clearing or limitations in the available wavelength coverage, which restrict the ability of the SED to constrain the innermost disc regions. In addition, factors such as grain growth, disc geometry, and optical depth effects can shift the effective sublimation front outward. Therefore, the derived $R_{\mathrm{in}}$ should be interpreted as an effective inner radius rather than a direct measurement of the physical dust sublimation boundary.

Panel (\emph{bottom left}) of Figure~\ref{fig:disc_model} displays the radial temperature profile of the disc midplane. The temperature decreases smoothly with increasing radius, following the expected power-law behavior for passive irradiated discs. No abrupt discontinuities are present, supporting the assumption of a continuous disc structure.

Panel (\emph{bottom right}) of Figure~\ref{fig:disc_model} illustrates the radial origin of the infrared emission, computed by integrating the contribution of each annulus to the total flux. The near-infrared emission is dominated by the innermost disc regions, while longer wavelengths arise from progressively larger radii. This radial decomposition demonstrates that the broadband SED effectively probes a wide range of disc scales, despite the limited wavelength coverage.

\begin{figure*}
\centering
 \includegraphics[width=1.\linewidth]{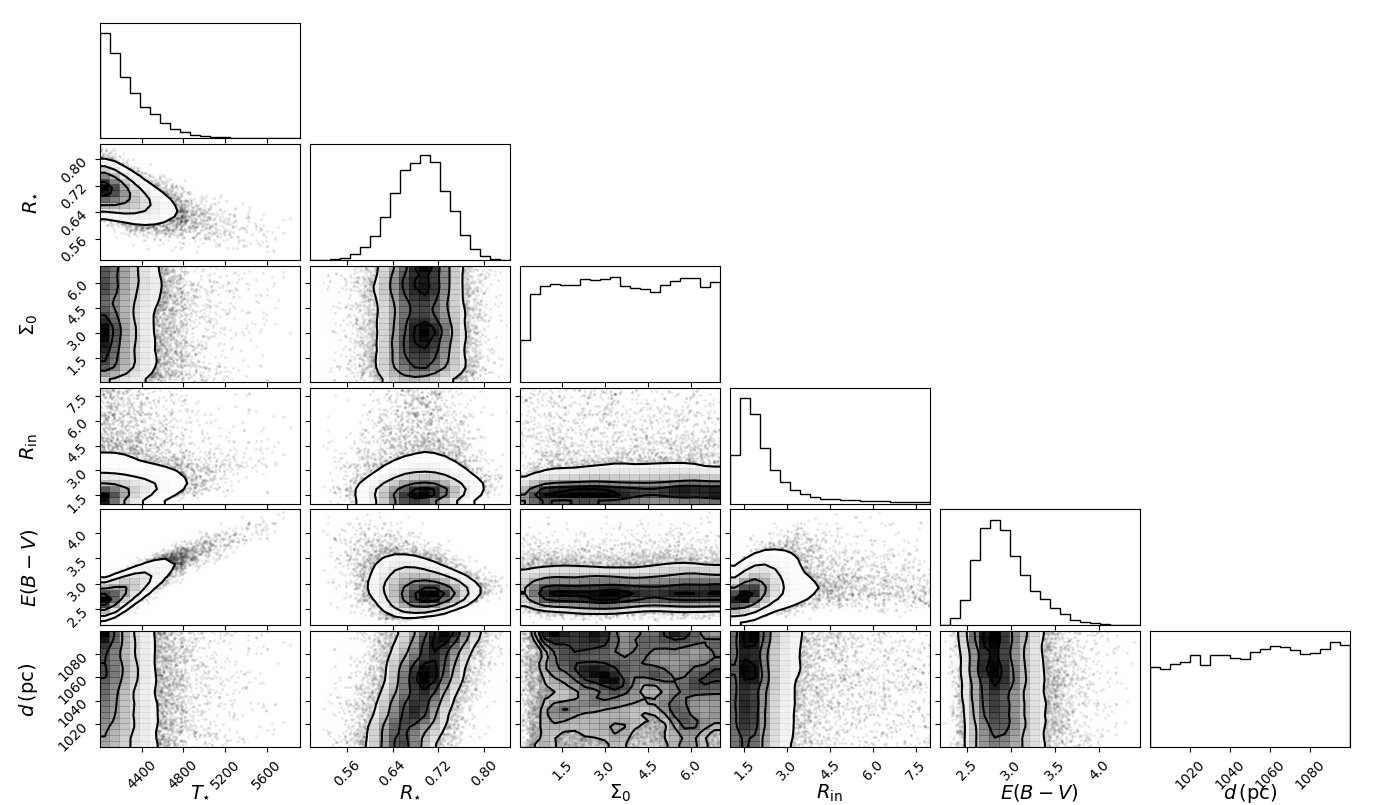}
\caption{
Posterior probability distributions of the free parameters obtained from the MCMC sampling. Contours correspond to 68\% and 95\% confidence levels.
}
\label{fig:corner}
\end{figure*}

The posterior probability distributions of the free parameters are shown in Figure~\ref{fig:corner}. The marginalized distributions are unimodal and well constrained, with finite covariances between correlated parameters such as $T_\star$ and $E(B-V)$. No evidence for multi-modal solutions is found within the explored parameter space, indicating that the adopted priors and wavelength coverage provide sufficient statistical leverage for stable inference.

The median values and 68\% credible intervals of the free parameters are summarized in Table~\ref{tab:sed_results}. The inferred stellar and disc properties are physically consistent and fall within the range expected for pre-main-sequence stars surrounded by circumstellar discs. Taken together, the SED fit, posterior distributions, and derived structural parameters support the interpretation of 2MASS J02512618+6012576 as a moderately embedded Class II YSO.

\begin{table}
\setlength{\tabcolsep}{5pt}
\renewcommand{\arraystretch}{1.5}
\centering
\caption{Posterior median values and 68\% credible intervals of the free parameters obtained from the MCMC SED fitting of 2MASS J02512618+6012576. The distance is constrained using a Gaussian prior from {\it Gaia} DR3.}
\begin{tabular}{lcc}
\hline
Parameter & Value & Unit \\
\hline
$T_\star$ & $4189^{+280}_{-140}$ & K \\
$R_\star$ & $0.711^{+0.042}_{-0.045}$ & $R_\odot$ \\
$R_{\mathrm{in}}$ & $1.92^{+0.60}_{-0.39}$ & AU \\
$\Sigma_0$ & $3.22^{+2.50}_{-2.10}$ & g\,cm$^{-2}$ \\
$E(B-V)$ & $2.98^{+0.34}_{-0.22}$ & mag \\
$d$ & $1052^{+33}_{-36}$ & pc \\
\hline
$M_{\mathrm{dust}}$ & $1.13^{+0.90}_{-0.73}\times10^{-3}$ & $M_\odot$ \\
\hline
\end{tabular}
\label{tab:sed_results}
\end{table}

\section{Discussion}

\subsection{Stellar Properties and Extinction}

The SED modelling yields an effective temperature of $T_\star \simeq 4200$\,K and a stellar radius of $R_\star \simeq 0.71\,R_\odot$. These values are consistent with a late-K-type pre-main-sequence object. The derived temperature is significantly lower than would be inferred from the observed {\it Gaia} colour index if interstellar reddening were neglected, confirming that extinction plays a dominant role in shaping the optical SED of the source.

The extinction is applied to the model fluxes using the relation given in Equation~(\ref{eq:Av}). The wavelength-dependent extinction is parameterised as

\begin{equation}
A_\lambda = k(\lambda, R_V) \, E(B-V),
\end{equation}
where $k(\lambda, R_{\rm V})$ is the extinction coefficient determined by the adopted extinction law and depends on the total-to-selective extinction ratio $R_{\rm V}$. In this work, $k(\lambda, R_{\rm V})$ is computed using the extinction law of \citet{Gordon2023}.

The inferred colour excess, $E(B-V) \approx 3$~mag, indicates substantial extinction along the line of sight. Given the location of the object within the IC~1848 region of the Soul Nebula, such high extinction is not implausible. However, the interpretation of this value requires caution. The conversion between $E(B-V)$ and {\it Gaia} colour excess depends sensitively on the adopted extinction law and on the assumed value of $R_{\rm V}$. In this study, a standard Galactic value of $R_{\rm V}=3.1$ was adopted. In dense star-forming environments, however, larger values of $R_{\rm V}$ are frequently reported. A different choice of $R_{\rm V}$ would modify the derived extinction and, consequently, the inferred stellar parameters.

To test the sensitivity of the inferred parameters to the adopted extinction law, the MCMC analysis was repeated with $R_{\rm V}=5$, representative of dense star-forming environments. The resulting posterior distributions remained broadly consistent with those obtained for $R_{\rm V}=3.1$. In particular, the inferred colour excess did not decrease significantly, with $E(B-V)=2.90^{+0.23}_{-0.17}$ mag for $R_{\rm V}=5$ compared to $E(B-V)=2.98^{+0.34}_{-0.22}$ mag for $R_{\rm V}=3.1$. This indicates that the relatively high extinction inferred for 2MASS J02512618+6012576 is not solely an artifact of adopting the standard Galactic extinction law, but is instead required by the broadband SED fit under the present modelling assumptions. The main effect of adopting a larger $R_V$ is a modest shift in the stellar parameters, yielding a slightly lower effective temperature ($\Delta T=-100$\,K) and a somewhat larger stellar radius ($\Delta R=0.07$\,$R_\odot$), while leaving the disc parameters and distance essentially unchanged.

Although the {\it Gaia} trigonometric parallax uncertainty is relatively large (about 25\%), incorporating it as a Gaussian prior in the MCMC analysis constrains the distance to $d \approx 1050$\,pc, consistent with the nominal Gaia value. This constraint significantly reduces the classical degeneracy between distance, stellar radius, and extinction that commonly affects broadband SED fitting.

\subsection{Disc Structure and Evolutionary Status}

The infrared excess clearly indicates the presence of a circumstellar disc. The modelling yields an inner disc radius of $R_{\mathrm{in}} \sim 2\,R_\star$ and a dust mass on the order of $10^{-3}\,M_\odot$. Such a disc mass is typical of Class II YSOs and is consistent with a relatively early evolutionary stage. The absence of extremely large inner cavities suggests that the object is unlikely to be a transition disc, although higher-resolution infrared data would be required to confirm this.

The disc surface density normalization remains moderately uncertain, reflecting the limited wavelength coverage and the absence of (sub)millimeter constraints. As a result, the total dust mass should be regarded as an order-of-magnitude estimate rather than a precise measurement.

\subsection{Inner disc Radius and Dust Sublimation}

The inferred inner disc radius, $R_{\mathrm{in}} \approx 2$ AU, is significantly larger than the expected dust sublimation radius for a $\sim4200$ K pre-main-sequence star. Adopting a characteristic dust sublimation temperature of $T_{\rm sub} \sim 1500$ K (e.g., \citealt{Pollack1994,Dullemond2001}) and using the approximation

\[
R_{\rm sub} \approx R_\star \left(\frac{T_\star}{T_{\rm sub}}\right)^2,
\]
the sublimation radius is estimated to be on the order of $0.03$–$0.05$ AU. The large discrepancy between $R_{\mathrm{in}}$ and $R_{\rm sub}$ suggests that the disc may possess a substantial inner cavity.

However, this interpretation must be treated with caution. The available photometric data extend only to 2.2\,$\mu$m, and the lack of mid-infrared constraints limits the sensitivity of the model to the detailed structure of the innermost disc regions. In addition, residual degeneracies between extinction, stellar luminosity, and disc geometry may artificially inflate the inferred inner radius. Higher-resolution infrared observations would be required to confirm whether the system exhibits genuine transitional disc characteristics.

\subsection{Multi-wavelength Image Predictions}

In addition to reproducing the SED, the model allows the generation of synthetic disc images at multiple wavelengths and viewing inclinations. These images demonstrate how the effective emitting region shifts outward with increasing wavelength and provide a spatial interpretation of the inferred disc structure.

At near-infrared wavelengths (e.g., $2~\mu$m), the emission is dominated by the stellar photosphere, with only a minor contribution from the innermost disc regions. At mid and far-infrared wavelengths ($\gtrsim10~\mu$m), disc emission becomes increasingly prominent, tracing progressively larger radii as cooler dust contributes to the observed flux.

\begin{figure}[ht]
    \centering
    \includegraphics[width=1.\linewidth]{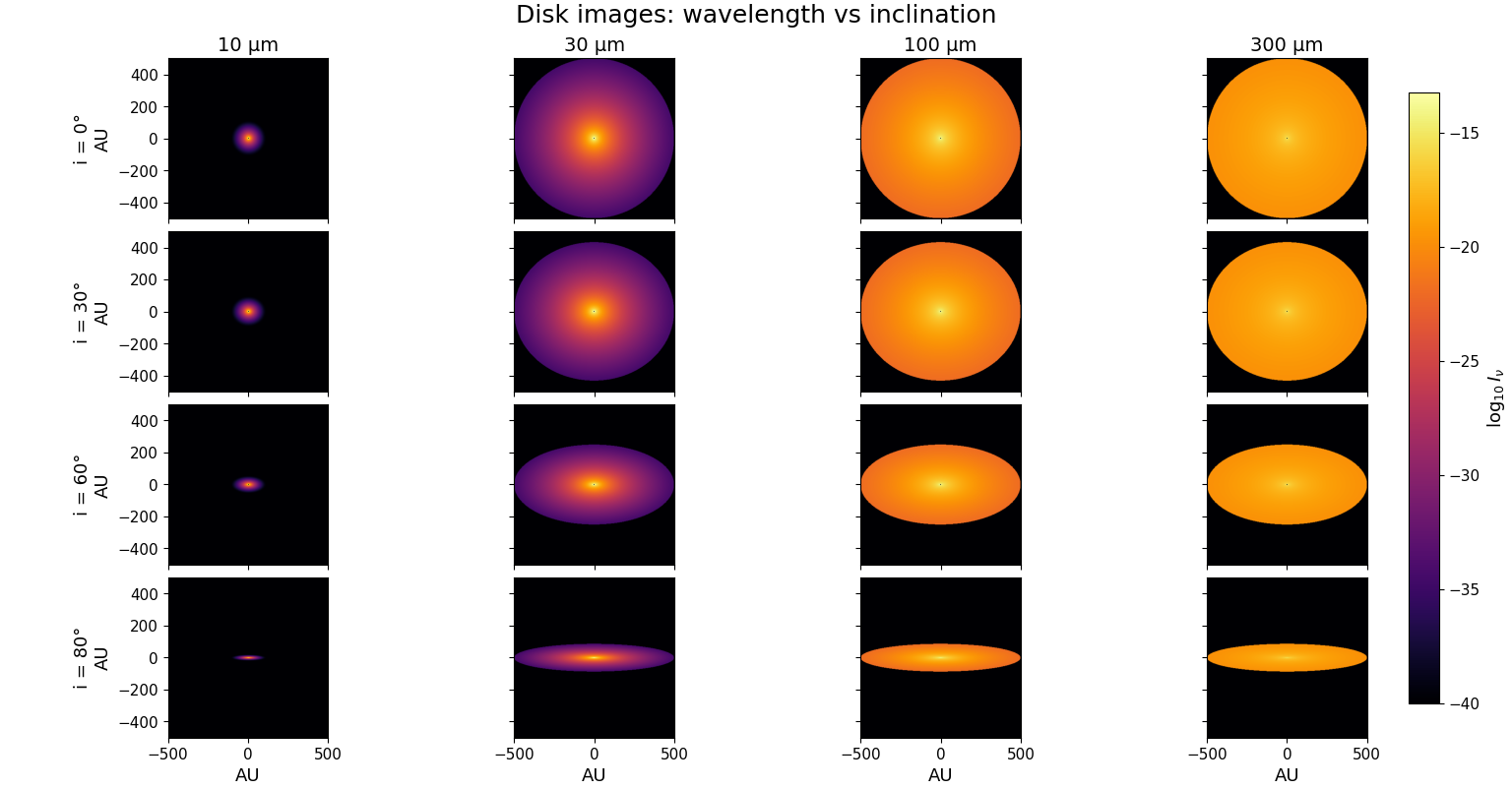}
    \caption{Synthetic disc images of 2MASS J02512618+6012576 at representative wavelengths and viewing inclinations derived from the best-fit model. The images are constructed by integrating the local thermal emission along the line of sight under the assumption of optically thin dust emission. Each panel is normalized to its peak intensity to highlight the spatial distribution of the emission. The apparent morphology and spatial extent vary significantly with wavelength, with shorter wavelengths tracing the inner, warmer regions and longer wavelengths probing cooler material at larger radii.}
    \label{fig:disc_images}
\end{figure}

Representative synthetic images of 2MASS J02512618+6012576 at selected wavelengths and inclinations are shown in Figure~\ref{fig:disc_images}. The images are computed by integrating the local emissivity along the line of sight, assuming optically thin dust emission, which provides a qualitative visualization of the spatial origin of the emission at different wavelengths. The panels in Figure~\ref{fig:disc_images} correspond to different viewing inclinations, illustrating projection effects on the morphology of the disc. The apparent morphology and surface brightness distribution vary significantly with wavelength, reflecting the radial temperature structure derived from the SED modelling.

\subsection{Evolutionary Classification}

Despite the relatively large inner disc radius inferred, the overall morphology of SED is consistent with that of a Class II YSO, in agreement with the classification of  \cite{Habali2026}. The source exhibits a clear infrared excess above the stellar photosphere, indicative of a substantial circumstellar disc. The derived dust mass of order $10^{-3}\,M_\odot$ is typical of classical T Tauri discs \citep[see][]{Andrews2005, Andrews2013} and does not suggest an advanced dispersal stage.

The apparent discrepancy between the inferred $R_{\mathrm{in}}$ and the expected dust sublimation radius may reflect limitations of the available wavelength coverage rather than the presence of a fully developed transitional cavity. In particular, the absence of mid-infrared and (sub)millimeter data restricts the ability of the model to accurately constrain the innermost disc structure. Within these limitations, the global SED shape supports a classification as a Class II pre-main-sequence object rather than a transition disc.

Therefore, 2MASS J02512618+6012576 is most consistently interpreted as a moderately embedded, late-type Class II YSO, although improved infrared observations would be required to definitively assess the presence or absence of an inner cavity.

\subsection{Degeneracies and Model Limitations}

Broadband SED fitting inherently suffers from parameter degeneracies, particularly between stellar temperature and extinction, and between stellar radius and distance. In the present analysis, the inclusion of a distance prior from Gaia reduces, but does not completely eliminate, these correlations. The posterior distributions exhibit elliptical covariances rather than open-ended degeneracies, indicating that the parameter space is statistically well behaved under the adopted assumptions.

Nevertheless, the derived extinction remains model-dependent. The use of a fixed extinction law ($R_{\rm V}=3.1$) and a single dust prescription may not fully capture the complexity of the local environment. To assess the sensitivity of the results to the adopted extinction law, we repeated the MCMC analysis with a higher value of $R_{\rm V}=5$, representative of dense star-forming environments. The resulting posterior distributions remain broadly consistent with those obtained for $R_{\rm V}=3.1$, and the inferred colour excess does not decrease significantly. This indicates that the relatively high extinction is not solely an artifact of assuming a standard Galactic extinction law, but is instead required by the SED fit under the current modelling framework.

Future spectroscopic observations or multi-band optical photometry would allow a more direct determination of the effective temperature and reddening, thereby breaking the residual degeneracy.

\subsection{Implications for Method Applicability}

This case study demonstrates that the modelling framework developed in this work can yield physically plausible stellar and disc parameters even for previously unstudied systems with limited prior information. The Bayesian approach, combined with external astrometric constraints, provides stable posterior solutions and realistic uncertainty estimates.

At the same time, the analysis highlights the sensitivity of broadband SED modelling to assumptions regarding extinction laws and distance constraints. For embedded YSOs, independent spectroscopic temperature estimates and improved extinction characterization are essential for fully disentangling stellar and circumstellar contributions.

Overall, 2MASS J02512618+6012576 is consistent with a moderately embedded, late-type pre-main-sequence star surrounded by a substantial circumstellar disc, supporting its classification as a bona fide YSO candidate.

\section{Conclusions}

In this study, we developed a Bayesian SED modelling framework designed to extract stellar and disc parameters from broadband photometric data. The method incorporates interstellar extinction, disc structure, and external distance constraints within a unified MCMC-based inference scheme.

As a demonstration case, the framework was applied to the previously uncharacterised YSO candidate 2MASS J02512618+6012576. Despite the limited wavelength coverage, the model successfully reproduces the observed SED and yields physically plausible stellar and disc parameters. The posterior distributions are well behaved and show no evidence of multi-modal degeneracies under the adopted priors.

The derived stellar properties are consistent with a late-type pre-main-sequence star, while the infrared excess and disc mass indicate the presence of a substantial circumstellar disc. Although the inferred inner disc radius exceeds the nominal dust sublimation radius, the current data do not provide sufficient leverage to confirm the presence of a transitional cavity. Within the limitations of the available photometry, the source is most consistently classified as a moderately embedded Class II YSO.

This application demonstrates that robust physical constraints can be derived from broadband SED fitting when extinction and distance are treated within a statistically consistent framework. Future extensions of the method to include mid-infrared and (sub)millimeter data will further improve constraints on disc structure and evolutionary state.

\textbf{ACKNOWLEDGEMENT}
This research has made use of the SIMBAD database, operated at CDS, Strasbourg, France. This research has made use of the VizieR catalogue access tool, CDS, Strasbourg, France (DOI: 10.26093/cds/vizier).

\bibliographystyle{mnras}
\bibliography{references}

@ARTICLE{Adams1987,
       author = {{Adams}, Fred C. and {Lada}, Charles J. and {Shu}, Frank H.},
        title = "{Spectral Evolution of Young Stellar Objects}",
      journal = {\apj},
         year = 1987,
        month = jan,
       volume = {312},
        pages = {788},
          doi = {10.1086/164924},
       adsurl = {https://ui.adsabs.harvard.edu/abs/1987ApJ...312..788A}
}

@ARTICLE{Andrews2005,
       author = {{Andrews}, Sean M. and {Williams}, Jonathan P.},
        title = "{Circumstellar Dust Disks in Taurus-Auriga: The Submillimeter Perspective}",
      journal = {\apj},
         year = 2005,
        month = oct,
       volume = {631},
       number = {2},
        pages = {1134-1160},
          doi = {10.1086/432712},
archivePrefix = {arXiv},
       eprint = {astro-ph/0506187},
 primaryClass = {astro-ph},
       adsurl = {https://ui.adsabs.harvard.edu/abs/2005ApJ...631.1134A}
}

@ARTICLE{Andrews2009,
       author = {{Andrews}, Sean M. and {Wilner}, D.~J. and {Hughes}, A.~M. and {Qi}, Chunhua and {Dullemond}, C.~P.},
        title = "{Protoplanetary Disk Structures in Ophiuchus}",
      journal = {\apj},
         year = 2009,
        month = aug,
       volume = {700},
       number = {2},
        pages = {1502-1523},
          doi = {10.1088/0004-637X/700/2/1502},
archivePrefix = {arXiv},
       eprint = {0906.0730},
 primaryClass = {astro-ph.EP},
       adsurl = {https://ui.adsabs.harvard.edu/abs/2009ApJ...700.1502A}
}

@ARTICLE{Andrews2013,
       author = {{Andrews}, Sean M. and {Rosenfeld}, Katherine A. and {Kraus}, Adam L. and {Wilner}, David J.},
        title = "{The Mass Dependence between Protoplanetary Disks and their Stellar Hosts}",
      journal = {\apj},
         year = 2013,
        month = jul,
       volume = {771},
       number = {2},
          eid = {129},
        pages = {129},
          doi = {10.1088/0004-637X/771/2/129},
archivePrefix = {arXiv},
       eprint = {1305.5262},
 primaryClass = {astro-ph.SR},
       adsurl = {https://ui.adsabs.harvard.edu/abs/2013ApJ...771..129A}
}

@ARTICLE{Bakis2022,
       author = {{Bak{\i}{\c{s}}}, V. and {Eker}, Z.},
        title = "{A Method of Improving Standard Stellar Luminosities with Multiband Standard Bolometric Corrections}",
      journal = {ActA},
         year = 2022,
        month = dec,
       volume = {72},
       number = {3},
        pages = {195-232},
          doi = {10.32023/0001-5237/72.3.4},
archivePrefix = {arXiv},
       eprint = {2208.04110},
 primaryClass = {astro-ph.SR},
       adsurl = {https://ui.adsabs.harvard.edu/abs/2022AcA....72..195B}
}

@ARTICLE{Beckwith1990,
       author = {{Beckwith}, Steven V.~W. and {Sargent}, Anneila I. and {Chini}, Rolf S. and {Guesten}, Rolf},
        title = "{A Survey for Circumstellar Disks around Young Stellar Objects}",
      journal = {\aj},
         year = 1990,
        month = mar,
       volume = {99},
        pages = {924},
          doi = {10.1086/115385},
       adsurl = {https://ui.adsabs.harvard.edu/abs/1990AJ.....99..924B}
}

@ARTICLE{Sphere2019,
       author = {{Beuzit}, J.-L. and {Vigan}, A. and {Mouillet}, D. and {Dohlen}, K. and {Gratton}, R. and {Boccaletti}, A. and {Sauvage}, J.-F. and {Schmid}, H.~M. and {Langlois}, M. and {Petit}, C. and {Baruffolo}, A. and {Feldt}, M. and {Milli}, J. and {Wahhaj}, Z. and {Abe}, L. and {Anselmi}, U. and {Antichi}, J. and {Barette}, R. and {Baudrand}, J. and {Baudoz}, P. and {Bazzon}, A. and {Bernardi}, P. and {Blanchard}, P. and {Brast}, R. and {Bruno}, P. and {Buey}, T. and {Carbillet}, M. and {Carle}, M. and {Cascone}, E. and {Chapron}, F. and {Charton}, J. and {Chauvin}, G. and {Claudi}, R. and {Costille}, A. and {De Caprio}, V. and {de Boer}, J. and {Delboulb{\'e}}, A. and {Desidera}, S. and {Dominik}, C. and {Downing}, M. and {Dupuis}, O. and {Fabron}, C. and {Fantinel}, D. and {Farisato}, G. and {Feautrier}, P. and {Fedrigo}, E. and {Fusco}, T. and {Gigan}, P. and {Ginski}, C. and {Girard}, J. and {Giro}, E. and {Gisler}, D. and {Gluck}, L. and {Gry}, C. and {Henning}, T. and {Hubin}, N. and {Hugot}, E. and {Incorvaia}, S. and {Jaquet}, M. and {Kasper}, M. and {Lagadec}, E. and {Lagrange}, A.-M. and {Le Coroller}, H. and {Le Mignant}, D. and {Le Ruyet}, B. and {Lessio}, G. and {Lizon}, J.-L. and {Llored}, M. and {Lundin}, L. and {Madec}, F. and {Magnard}, Y. and {Marteaud}, M. and {Martinez}, P. and {Maurel}, D. and {M{\'e}nard}, F. and {Mesa}, D. and {M{\"o}ller-Nilsson}, O. and {Moulin}, T. and {Moutou}, C. and {Orign{\'e}}, A. and {Parisot}, J. and {Pavlov}, A. and {Perret}, D. and {Pragt}, J. and {Puget}, P. and {Rabou}, P. and {Ramos}, J. and {Reess}, J.-M. and {Rigal}, F. and {Rochat}, S. and {Roelfsema}, R. and {Rousset}, G. and {Roux}, A. and {Saisse}, M. and {Salasnich}, B. and {Santambrogio}, E. and {Scuderi}, S. and {Segransan}, D. and {Sevin}, A. and {Siebenmorgen}, R. and {Soenke}, C. and {Stadler}, E. and {Suarez}, M. and {Tiph{\`e}ne}, D. and {Turatto}, M. and {Udry}, S. and {Vakili}, F. and {Waters}, L.~B.~F.~M. and {Weber}, L. and {Wildi}, F. and {Zins}, G. and {Zurlo}, A.},
        title = "{SPHERE: the exoplanet imager for the Very Large Telescope}",
      journal = {\aap},
         year = 2019,
        month = nov,
       volume = {631},
          eid = {A155},
        pages = {A155},
          doi = {10.1051/0004-6361/201935251},
archivePrefix = {arXiv},
       eprint = {1902.04080},
 primaryClass = {astro-ph.IM},
       adsurl = {https://ui.adsabs.harvard.edu/abs/2019A&A...631A.155B}
}

@ARTICLE{Chambers2016,
       author = {{Chambers}, K.~C. and {Magnier}, E.~A. and {Metcalfe}, N. and {Flewelling}, H.~A. and {Huber}, M.~E. and {Waters}, C.~Z. and {Denneau}, L. and {Draper}, P.~W. and {Farrow}, D. and {Finkbeiner}, D.~P. and {Holmberg}, C. and {Koppenhoefer}, J. and {Price}, P.~A. and {Rest}, A. and {Saglia}, R.~P. and {Schlafly}, E.~F. and {Smartt}, S.~J. and {Sweeney}, W. and {Wainscoat}, R.~J. and {Burgett}, W.~S. and {Chastel}, S. and {Grav}, T. and {Heasley}, J.~N. and {Hodapp}, K.~W. and {Jedicke}, R. and {Kaiser}, N. and {Kudritzki}, R.-P. and {Luppino}, G.~A. and {Lupton}, R.~H. and {Monet}, D.~G. and {Morgan}, J.~S. and {Onaka}, P.~M. and {Shiao}, B. and {Stubbs}, C.~W. and {Tonry}, J.~L. and {White}, R. and {Ba{\~n}ados}, E. and {Bell}, E.~F. and {Bender}, R. and {Bernard}, E.~J. and {Boegner}, M. and {Boffi}, F. and {Botticella}, M.~T. and {Calamida}, A. and {Casertano}, S. and {Chen}, W.-P. and {Chen}, X. and {Cole}, S. and {Deacon}, N. and {Frenk}, C. and {Fitzsimmons}, A. and {Gezari}, S. and {Gibbs}, V. and {Goessl}, C. and {Goggia}, T. and {Gourgue}, R. and {Goldman}, B. and {Grant}, P. and {Grebel}, E.~K. and {Hambly}, N.~C. and {Hasinger}, G. and {Heavens}, A.~F. and {Heckman}, T.~M. and {Henderson}, R. and {Henning}, T. and {Holman}, M. and {Hopp}, U. and {Ip}, W.-H. and {Isani}, S. and {Jackson}, M. and {Keyes}, C.~D. and {Koekemoer}, A.~M. and {Kotak}, R. and {Le}, D. and {Liska}, D. and {Long}, K.~S. and {Lucey}, J.~R. and {Liu}, M. and {Martin}, N.~F. and {Masci}, G. and {McLean}, B. and {Mindel}, E. and {Misra}, P. and {Morganson}, E. and {Murphy}, D.~N.~A. and {Obaika}, A. and {Narayan}, G. and {Nieto-Santisteban}, M.~A. and {Norberg}, P. and {Peacock}, J.~A. and {Pier}, E.~A. and {Postman}, M. and {Primak}, N. and {Rae}, C. and {Rai}, A. and {Riess}, A. and {Riffeser}, A. and {Rix}, H.~W. and {R{\"o}ser}, S. and {Russel}, R. and {Rutz}, L. and {Schilbach}, E. and {Schultz}, A.~S.~B. and {Scolnic}, D. and {Strolger}, L. and {Szalay}, A. and {Seitz}, S. and {Small}, E. and {Smith}, K.~W. and {Soderblom}, D.~R. and {Taylor}, P. and {Thomson}, R. and {Taylor}, A.~N. and {Thakar}, A.~R. and {Thiel}, J. and {Thilker}, D. and {Unger}, D. and {Urata}, Y. and {Valenti}, J. and {Wagner}, J. and {Walder}, T. and {Walter}, F. and {Watters}, S.~P. and {Werner}, S. and {Wood-Vasey}, W.~M. and {Wyse}, R.},
        title = "{The Pan-STARRS1 Surveys}",
      journal = {arXiv e-prints},
         year = 2016,
        month = dec,
          eid = {arXiv:1612.05560},
        pages = {arXiv:1612.05560},
          doi = {10.48550/arXiv.1612.05560},
archivePrefix = {arXiv},
       eprint = {1612.05560},
 primaryClass = {astro-ph.IM},
       adsurl = {https://ui.adsabs.harvard.edu/abs/2016arXiv161205560C}
}

@ARTICLE{Chiang1997,
       author = {{Chiang}, E.~I. and {Goldreich}, P.},
        title = "{Spectral Energy Distributions of T Tauri Stars with Passive Circumstellar Disks}",
      journal = {ApJ},
         year = 1997,
        month = nov,
       volume = {490},
       number = {1},
        pages = {368-376},
          doi = {10.1086/304869},
archivePrefix = {arXiv},
       eprint = {astro-ph/9706042},
 primaryClass = {astro-ph},
       adsurl = {https://ui.adsabs.harvard.edu/abs/1997ApJ...490..368C}
}

@ARTICLE{Dullemond2001,
       author = {{Dullemond}, C.~P. and {Dominik}, C. and {Natta}, A.},
        title = "{Passive Irradiated Circumstellar Disks with an Inner Hole}",
      journal = {\apj},
         year = 2001,
        month = oct,
       volume = {560},
       number = {2},
        pages = {957-969},
          doi = {10.1086/323057},
archivePrefix = {arXiv},
       eprint = {astro-ph/0106470},
 primaryClass = {astro-ph},
       adsurl = {https://ui.adsabs.harvard.edu/abs/2001ApJ...560..957D}
}

@software{Dullemond2012,
       author = {{Dullemond}, C.~P. and {Juhasz}, A. and {Pohl}, A. and {Sereshti}, F. and {Shetty}, R. and {Peters}, T. and {Commercon}, B. and {Flock}, M.},
        title = "{RADMC-3D: A multi-purpose radiative transfer tool}",
 howpublished = {Astrophysics Source Code Library, record ascl:1202.015},
         year = 2012,
        month = feb,
          eid = {ascl:1202.015},
archivePrefix = {ascl},
       eprint = {1202.015},
       adsurl = {https://ui.adsabs.harvard.edu/abs/2012ascl.soft02015D}
}

@ARTICLE{ForemanMackey2013,
       author = {{Foreman-Mackey}, Daniel and {Hogg}, David W. and {Lang}, Dustin and {Goodman}, Jonathan},
        title = "{emcee: The MCMC Hammer}",
      journal = {\pasp},
         year = 2013,
        month = mar,
       volume = {125},
       number = {925},
        pages = {306},
          doi = {10.1086/670067},
archivePrefix = {arXiv},
       eprint = {1202.3665},
 primaryClass = {astro-ph.IM},
       adsurl = {https://ui.adsabs.harvard.edu/abs/2013PASP..125..306F}
}

@ARTICLE{GaiaDR3,
       author = {{Gaia Collaboration} and {Vallenari}, A. and {Brown}, A.~G.~A. and {Prusti}, T. and {de Bruijne}, J.~H.~J. and {Arenou}, F. and {Babusiaux}, C. and {Biermann}, M. and {Creevey}, O.~L. and {Ducourant}, C. and {Evans}, D.~W. and {Eyer}, L. and {Guerra}, R. and {Hutton}, A. and {Jordi}, C. and {Klioner}, S.~A. and {Lammers}, U.~L. and {Lindegren}, L. and {Luri}, X. and {Mignard}, F. and {Panem}, C. and {Pourbaix}, D. and {Randich}, S. and {Sartoretti}, P. and {Soubiran}, C. and {Tanga}, P. and {Walton}, N.~A. and {Bailer-Jones}, C.~A.~L. and {Bastian}, U. and {Drimmel}, R. and {Jansen}, F. and {Katz}, D. and {Lattanzi}, M.~G. and {van Leeuwen}, F. and {Bakker}, J. and {Cacciari}, C. and {Casta{\~n}eda}, J. and {De Angeli}, F. and {Fabricius}, C. and {Fouesneau}, M. and {Fr{\'e}mat}, Y. and {Galluccio}, L. and {Guerrier}, A. and {Heiter}, U. and {Masana}, E. and {Messineo}, R. and {Mowlavi}, N. and {Nicolas}, C. and {Nienartowicz}, K. and {Pailler}, F. and {Panuzzo}, P. and {Riclet}, F. and {Roux}, W. and {Seabroke}, G.~M. and {Sordo}, R. and {Th{\'e}venin}, F. and {Gracia-Abril}, G. and {Portell}, J. and {Teyssier}, D. and {Altmann}, M. and {Andrae}, R. and {Audard}, M. and {Bellas-Velidis}, I. and {Benson}, K. and {Berthier}, J. and {Blomme}, R. and {Burgess}, P.~W. and {Busonero}, D. and {Busso}, G. and {C{\'a}novas}, H. and {Carry}, B. and {Cellino}, A. and {Cheek}, N. and {Clementini}, G. and {Damerdji}, Y. and {Davidson}, M. and {de Teodoro}, P. and {Nu{\~n}ez Campos}, M. and {Delchambre}, L. and {Dell'Oro}, A. and {Esquej}, P. and {Fern{\'a}ndez-Hern{\'a}ndez}, J. and {Fraile}, E. and {Garabato}, D. and {Garc{\'\i}a-Lario}, P. and {Gosset}, E. and {Haigron}, R. and {Halbwachs}, J.-L. and {Hambly}, N.~C. and {Harrison}, D.~L. and {Hern{\'a}ndez}, J. and {Hestroffer}, D. and {Hodgkin}, S.~T. and {Holl}, B. and {Jan{\ss}en}, K. and {Jevardat de Fombelle}, G. and {Jordan}, S. and {Krone-Martins}, A. and {Lanzafame}, A.~C. and {L{\"o}ffler}, W. and {Marchal}, O. and {Marrese}, P.~M. and {Moitinho}, A. and {Muinonen}, K. and {Osborne}, P. and {Pancino}, E. and {Pauwels}, T. and {Recio-Blanco}, A. and {Reyl{\'e}}, C. and {Riello}, M. and {Rimoldini}, L. and {Roegiers}, T. and {Rybizki}, J. and {Sarro}, L.~M. and {Siopis}, C. and {Smith}, M. and {Sozzetti}, A. and {Utrilla}, E. and {van Leeuwen}, M. and {Abbas}, U. and {{\'A}brah{\'a}m}, P. and {Abreu Aramburu}, A. and {Aerts}, C. and {Aguado}, J.~J. and {Ajaj}, M. and {Aldea-Montero}, F. and {Altavilla}, G. and {{\'A}lvarez}, M.~A. and {Alves}, J. and {Anders}, F. and {Anderson}, R.~I. and {Anglada Varela}, E. and {Antoja}, T. and {Baines}, D. and {Baker}, S.~G. and {Balaguer-N{\'u}{\~n}ez}, L. and {Balbinot}, E. and {Balog}, Z. and {Barache}, C. and {Barbato}, D. and {Barros}, M. and {Barstow}, M.~A. and {Bartolom{\'e}}, S. and {Bassilana}, J.-L. and {Bauchet}, N. and {Becciani}, U. and {Bellazzini}, M. and {Berihuete}, A. and {Bernet}, M. and {Bertone}, S. and {Bianchi}, L. and {Binnenfeld}, A. and {Blanco-Cuaresma}, S. and {Blazere}, A. and {Boch}, T. and {Bombrun}, A. and {Bossini}, D. and {Bouquillon}, S. and {Bragaglia}, A. and {Bramante}, L. and {Breedt}, E. and {Bressan}, A. and {Brouillet}, N. and {Brugaletta}, E. and {Bucciarelli}, B. and {Burlacu}, A. and {Butkevich}, A.~G. and {Buzzi}, R. and {Caffau}, E. and {Cancelliere}, R. and {Cantat-Gaudin}, T. and {Carballo}, R. and {Carlucci}, T. and {Carnerero}, M.~I. and {Carrasco}, J.~M. and {Casamiquela}, L. and {Castellani}, M. and {Castro-Ginard}, A. and {Chaoul}, L. and {Charlot}, P. and {Chemin}, L. and {Chiaramida}, V. and {Chiavassa}, A. and {Chornay}, N. and {Comoretto}, G. and {Contursi}, G. and {Cooper}, W.~J. and {Cornez}, T. and {Cowell}, S. and {Crifo}, F. and {Cropper}, M. and {Crosta}, M. and {Crowley}, C. and {Dafonte}, C. and {Dapergolas}, A. and {David}, M. and {David}, P. and {de Laverny}, P. and {De Luise}, F. and {De March}, R.},
        title = "{Gaia Data Release 3. Summary of the content and survey properties}",
      journal = {\aap},
         year = 2023,
        month = jun,
       volume = {674},
          eid = {A1},
        pages = {A1},
          doi = {10.1051/0004-6361/202243940},
archivePrefix = {arXiv},
       eprint = {2208.00211},
 primaryClass = {astro-ph.GA},
       adsurl = {https://ui.adsabs.harvard.edu/abs/2023A&A...674A...1G}
}

@ARTICLE{Gordon2023,
       author = {{Gordon}, Karl D. and {Clayton}, Geoffrey C. and {Decleir}, Marjorie and {Fitzpatrick}, E.~L. and {Massa}, Derck and {Misselt}, Karl A. and {Tollerud}, Erik J.},
        title = "{One Relation for All Wavelengths: The Far-ultraviolet to Mid-infrared Milky Way Spectroscopic R(V)-dependent Dust Extinction Relationship}",
      journal = {\apj},
         year = 2023,
        month = jun,
       volume = {950},
       number = {2},
          eid = {86},
        pages = {86},
          doi = {10.3847/1538-4357/accb59},
archivePrefix = {arXiv},
       eprint = {2304.01991},
 primaryClass = {astro-ph.GA},
       adsurl = {https://ui.adsabs.harvard.edu/abs/2023ApJ...950...86G}
}

@ARTICLE{Habali2026,
       author = {{Habal{\i}}, Ay{\textcommabelow s}e Yadikar and {Bak{\i}{\textcommabelow s}}, Volkan},
        title = "{A color--magnitude approach to YSO classification using SED slope and Gaia-distance-calibrated WISE/2MASS photometry}",
      journal = {Advances in Space Research},
         year = 2026,
        month = feb,
       volume = {77},
       number = {3},
        pages = {4018-4039},
          doi = {10.1016/j.asr.2025.11.050},
       adsurl = {https://ui.adsabs.harvard.edu/abs/2026AdSpR..77.4018H}
}

@ARTICLE{Hildebrand1983,
       author = {{Hildebrand}, R.~H.},
        title = "{The determination of cloud masses and dust characteristics from submillimetre thermal emission.}",
      journal = {\qjras},
         year = 1983,
        month = sep,
       volume = {24},
        pages = {267-282},
       adsurl = {https://ui.adsabs.harvard.edu/abs/1983QJRAS..24..267H}
}

@ARTICLE{Hughes2017,
       author = {{Hughes}, A. Meredith and {Lieman-Sifry}, Jesse and {Flaherty}, Kevin M. and {Daley}, Cail M. and {Roberge}, Aki and {K{\'o}sp{\'a}l}, {\'A}gnes and {Mo{\'o}r}, Attila and {Kamp}, Inga and {Wilner}, David J. and {Andrews}, Sean M. and {Kastner}, Joel H. and {{\'A}brah{\'a}m}, Peter},
        title = "{Radial Surface Density Profiles of Gas and Dust in the Debris Disk around 49 Ceti}",
      journal = {\apj},
         year = 2017,
        month = apr,
       volume = {839},
       number = {2},
          eid = {86},
        pages = {86},
          doi = {10.3847/1538-4357/aa6b04},
archivePrefix = {arXiv},
       eprint = {1704.01972},
 primaryClass = {astro-ph.EP},
       adsurl = {https://ui.adsabs.harvard.edu/abs/2017ApJ...839...86H}
}

@ARTICLE{Kobayashi2011,
       author = {{Kobayashi}, Hiroshi and {Kimura}, Hiroshi and {Watanabe}, Sei-ichiro and {Yamamoto}, Tetsuo and {M{\"u}ller}, Sebastian},
        title = "{Sublimation temperature of circumstellar dust particles and its importance for dust ring formation}",
      journal = {Earth, Planets and Space},
         year = 2011,
        month = oct,
       volume = {63},
       number = {10},
        pages = {1067-1075},
          doi = {10.5047/eps.2011.03.012},
archivePrefix = {arXiv},
       eprint = {1104.5627},
 primaryClass = {astro-ph.EP},
       adsurl = {https://ui.adsabs.harvard.edu/abs/2011EP&S...63.1067K}
}

@INPROCEEDINGS{Lada1987,
       author = {{Lada}, Charles J.},
        title = "{Star formation: from OB associations to protostars.}",
    booktitle = {Star Forming Regions},
         year = 1987,
       editor = {{Peimbert}, Manuel and {Jugaku}, Jun},
       series = {IAU Symposium},
       volume = {115},
        month = jan,
        pages = {1},
       adsurl = {https://ui.adsabs.harvard.edu/abs/1987IAUS..115....1L}
}

@ARTICLE{LyndenBell1974,
       author = {{Lynden-Bell}, D. and {Pringle}, J.~E.},
        title = "{The evolution of viscous discs and the origin of the nebular variables.}",
      journal = {\mnras},
         year = 1974,
        month = sep,
       volume = {168},
        pages = {603-637},
          doi = {10.1093/mnras/168.3.603},
       adsurl = {https://ui.adsabs.harvard.edu/abs/1974MNRAS.168..603L}
}

@ARTICLE{Neugebauer1984,
       author = {{Neugebauer}, G. and {Soifer}, B.~T. and {Miley}, G. and {Young}, E. and {Beichman}, C.~A. and {Clegg}, P.~E. and {Habing}, H.~J. and {Harris}, S. and {Low}, F.~J. and {Rowan-Robinson}, M.},
        title = "{IRAS observations of radio-quiet and radio-loud quasars.}",
      journal = {\apjl},
         year = 1984,
        month = mar,
       volume = {278},
        pages = {L83-L85},
          doi = {10.1086/184229},
       adsurl = {https://ui.adsabs.harvard.edu/abs/1984ApJ...278L..83N}
}

@ARTICLE{Nhung2017,
       author = {{Nhung}, P.~T. and {Hoai}, D.~T. and {Tuan-Anh}, P. and {Diep}, P.~N. and {Phuong}, N.~T. and {Thao}, N.~T. and {Darriulat}, P.},
        title = "{High-resolution ALMA observation of the $^{12}$CO(3-2) and 350 GHz continuum emissions of the debris disc of 49 Ceti}",
      journal = {MNRAS},
         year = 2017,
        month = aug,
       volume = {469},
       number = {4},
        pages = {4726-4739},
          doi = {10.1093/mnras/stx1125},
archivePrefix = {arXiv},
       eprint = {1701.02131},
 primaryClass = {astro-ph.SR},
       adsurl = {https://ui.adsabs.harvard.edu/abs/2017MNRAS.469.4726N}
}

@ARTICLE{Herschel2010,
       author = {{Pilbratt}, G.~L. and {Riedinger}, J.~R. and {Passvogel}, T. and {Crone}, G. and {Doyle}, D. and {Gageur}, U. and {Heras}, A.~M. and {Jewell}, C. and {Metcalfe}, L. and {Ott}, S. and {Schmidt}, M.},
        title = "{Herschel Space Observatory. An ESA facility for far-infrared and submillimetre astronomy}",
      journal = {\aap},
         year = 2010,
        month = jul,
       volume = {518},
          eid = {L1},
        pages = {L1},
          doi = {10.1051/0004-6361/201014759},
archivePrefix = {arXiv},
       eprint = {1005.5331},
 primaryClass = {astro-ph.IM},
       adsurl = {https://ui.adsabs.harvard.edu/abs/2010A&A...518L...1P}
}

@ARTICLE{Pinte2006,
       author = {{Pinte}, C. and {M{\'e}nard}, F. and {Duch{\^e}ne}, G. and {Bastien}, P.},
        title = "{Monte Carlo radiative transfer in protoplanetary disks}",
      journal = {A\&A},
         year = 2006,
        month = dec,
       volume = {459},
       number = {3},
        pages = {797-804},
          doi = {10.1051/0004-6361:20053275},
archivePrefix = {arXiv},
       eprint = {astro-ph/0606550},
 primaryClass = {astro-ph},
       adsurl = {https://ui.adsabs.harvard.edu/abs/2006A&A...459..797P}
}

@ARTICLE{Pollack1994,
       author = {{Pollack}, James B. and {Hollenbach}, David and {Beckwith}, Steven and {Simonelli}, Damon P. and {Roush}, Ted and {Fong}, Wesley},
        title = "{Composition and Radiative Properties of Grains in Molecular Clouds and Accretion Disks}",
      journal = {\apj},
         year = 1994,
        month = feb,
       volume = {421},
        pages = {615},
          doi = {10.1086/173677},
       adsurl = {https://ui.adsabs.harvard.edu/abs/1994ApJ...421..615P}
}

@ARTICLE{Sharpless1955,
       author = {{Sharpless}, Stewart},
        title = "{The distances and dimensions of IC 1805, IC 1848, and IC 410.}",
      journal = {AJ},
         year = 1955,
        month = jun,
       volume = {60},
        pages = {178-179},
          doi = {10.1086/107207},
       adsurl = {https://ui.adsabs.harvard.edu/abs/1955AJ.....60..178S}
}

@ARTICLE{Shu1977,
       author = {{Shu}, F.~H.},
        title = "{Self-similar collapse of isothermal spheres and star formation.}",
      journal = {\apj},
         year = 1977,
        month = jun,
       volume = {214},
        pages = {488-497},
          doi = {10.1086/155274},
       adsurl = {https://ui.adsabs.harvard.edu/abs/1977ApJ...214..488S}
}

@ARTICLE{Skrutskie2006,
       author = {{Skrutskie}, M.~F. and {Cutri}, R.~M. and {Stiening}, R. and {Weinberg}, M.~D. and {Schneider}, S. and {Carpenter}, J.~M. and {Beichman}, C. and {Capps}, R. and {Chester}, T. and {Elias}, J. and {Huchra}, J. and {Liebert}, J. and {Lonsdale}, C. and {Monet}, D.~G. and {Price}, S. and {Seitzer}, P. and {Jarrett}, T. and {Kirkpatrick}, J.~D. and {Gizis}, J.~E. and {Howard}, E. and {Evans}, T. and {Fowler}, J. and {Fullmer}, L. and {Hurt}, R. and {Light}, R. and {Kopan}, E.~L. and {Marsh}, K.~A. and {McCallon}, H.~L. and {Tam}, R. and {Van Dyk}, S. and {Wheelock}, S.},
        title = "{The Two Micron All Sky Survey (2MASS)}",
      journal = {\aj},
         year = 2006,
        month = feb,
       volume = {131},
       number = {2},
        pages = {1163-1183},
          doi = {10.1086/498708},
       adsurl = {https://ui.adsabs.harvard.edu/abs/2006AJ....131.1163S}
}

@ARTICLE{Virtanen2020,
       author = {{Virtanen}, Pauli and {Gommers}, Ralf and {Oliphant}, Travis E. and {Haberland}, Matt and {Reddy}, Tyler and {Cournapeau}, David and {Burovski}, Evgeni and {Peterson}, Pearu and {Weckesser}, Warren and {Bright}, Jonathan and {van der Walt}, St{\'e}fan J. and {Brett}, Matthew and {Wilson}, Joshua and {Millman}, K. Jarrod and {Mayorov}, Nikolay and {Nelson}, Andrew R.~J. and {Jones}, Eric and {Kern}, Robert and {Larson}, Eric and {Carey}, C.~J. and {Polat}, {\.I}lhan and {Feng}, Yu and {Moore}, Eric W. and {VanderPlas}, Jake and {Laxalde}, Denis and {Perktold}, Josef and {Cimrman}, Robert and {Henriksen}, Ian and {Quintero}, E.~A. and {Harris}, Charles R. and {Archibald}, Anne M. and {Ribeiro}, Ant{\^o}nio H. and {Pedregosa}, Fabian and {van Mulbregt}, Paul and {SciPy 1.  0 Contributors}},
        title = "{SciPy 1.0: fundamental algorithms for scientific computing in Python}",
      journal = {Nature Medicine},
         year = 2020,
        month = feb,
       volume = {17},
        pages = {261-272},
          doi = {10.1038/s41592-019-0686-2},
archivePrefix = {arXiv},
       eprint = {1907.10121},
 primaryClass = {cs.MS},
       adsurl = {https://ui.adsabs.harvard.edu/abs/2020NatMe..17..261V}
}

@ARTICLE{Spitzer2004,
       author = {{Werner}, M.~W. and {Roellig}, T.~L. and {Low}, F.~J. and {Rieke}, G.~H. and {Rieke}, M. and {Hoffmann}, W.~F. and {Young}, E. and {Houck}, J.~R. and {Brandl}, B. and {Fazio}, G.~G. and {Hora}, J.~L. and {Gehrz}, R.~D. and {Helou}, G. and {Soifer}, B.~T. and {Stauffer}, J. and {Keene}, J. and {Eisenhardt}, P. and {Gallagher}, D. and {Gautier}, T.~N. and {Irace}, W. and {Lawrence}, C.~R. and {Simmons}, L. and {Van Cleve}, J.~E. and {Jura}, M. and {Wright}, E.~L. and {Cruikshank}, D.~P.},
        title = "{The Spitzer Space Telescope Mission}",
      journal = {\apjs},
         year = 2004,
        month = sep,
       volume = {154},
       number = {1},
        pages = {1-9},
          doi = {10.1086/422992},
archivePrefix = {arXiv},
       eprint = {astro-ph/0406223},
 primaryClass = {astro-ph},
       adsurl = {https://ui.adsabs.harvard.edu/abs/2004ApJS..154....1W}
}

@ARTICLE{Alma2009,
       author = {{Wootten}, Alwyn and {Thompson}, A. Richard},
        title = "{The Atacama Large Millimeter/Submillimeter Array}",
      journal = {IEEE Proceedings},
         year = 2009,
        month = aug,
       volume = {97},
       number = {8},
        pages = {1463-1471},
          doi = {10.1109/JPROC.2009.2020572},
archivePrefix = {arXiv},
       eprint = {0904.3739},
 primaryClass = {astro-ph.IM},
       adsurl = {https://ui.adsabs.harvard.edu/abs/2009IEEEP..97.1463W}
}

@ARTICLE{Wright2010,
       author = {{Wright}, Edward L. and {Eisenhardt}, Peter R.~M. and {Mainzer}, Amy K. and {Ressler}, Michael E. and {Cutri}, Roc M. and {Jarrett}, Thomas and {Kirkpatrick}, J. Davy and {Padgett}, Deborah and {McMillan}, Robert S. and {Skrutskie}, Michael and {Stanford}, S.~A. and {Cohen}, Martin and {Walker}, Russell G. and {Mather}, John C. and {Leisawitz}, David and {Gautier}, III, Thomas N. and {McLean}, Ian and {Benford}, Dominic and {Lonsdale}, Carol J. and {Blain}, Andrew and {Mendez}, Bryan and {Irace}, William R. and {Duval}, Valerie and {Liu}, Fengchuan and {Royer}, Don and {Heinrichsen}, Ingolf and {Howard}, Joan and {Shannon}, Mark and {Kendall}, Martha and {Walsh}, Amy L. and {Larsen}, Mark and {Cardon}, Joel G. and {Schick}, Scott and {Schwalm}, Mark and {Abid}, Mohamed and {Fabinsky}, Beth and {Naes}, Larry and {Tsai}, Chao-Wei},
        title = "{The Wide-field Infrared Survey Explorer (WISE): Mission Description and Initial On-orbit Performance}",
      journal = {\aj},
         year = 2010,
        month = dec,
       volume = {140},
       number = {6},
        pages = {1868-1881},
          doi = {10.1088/0004-6256/140/6/1868},
archivePrefix = {arXiv},
       eprint = {1008.0031},
 primaryClass = {astro-ph.IM},
       adsurl = {https://ui.adsabs.harvard.edu/abs/2010AJ....140.1868W}
}

\end{document}